\documentclass{article}
\usepackage{romp}
\usepackage{graphicx}

\title{Application of Stochastic Variational Method to Hydrodynamics}
\author{T. Koide \\ Instituto de F\'{\i}sica, Universidade Federal do Rio de Janeiro,\\ C.P.
68528, 21941-972, Rio de Janeiro, Brazil \\ (e-mail: tomoikoide@gmail.com)\\ 
}

\begin{document}
\maketitle

\begin{abstract}
We apply the stochastic variational method to the action of the ideal fluid and showed that the Navier-Stokes equation is derived. 
In this variational method, the effect of dissipation is realized as the direct consequence of the 
fluctuation dissipation theorem. Differently from the previous works \cite{kk1,kk2}, we parameterize the Lagrangian of SVM in more general form. The form of the obtained equation is not modified but the definition of the transport coefficients are changed.
We further discuss the formulation of SVM using the Hamiltonian and show that 
the variation of the Hamiltonian gives the same result as the case of the Lagrangian.
\end{abstract}


\section{Introduction : origin of stochasticity in variational approach}

The variational principle is one of the important guiding principles in physics.
The usual variational principle, however, is not applicable when
irreversible dynamics is present. It is because dissipation is caused by energy
exchange processes between macroscopic and microscopic motions whose information is not included in Lagrangian. 
For example, the Euler equation can be derived from 
the Lagrangian ${\cal L}(\varepsilon, \rho^m, {\bf v})$ 
which is a function only of hydrodynamic variables: the energy density $\varepsilon$, 
the mass density $\rho^m$ and the fluid velocity ${\bf v}$. 
On the other hand, the viscosity of fluids is induced by the collisions between the constituent molecules of fluids. Thus, to introduce the effect of viscosity in the framework of the variational method, 
a new term, which represents the microscopic degree of freedom (molecules), is added to the original Lagrangian. 
This approach is known as Rayleigh's dissipation function method, 
but it is not easy to specify uniquely this additional term.

\begin{figure}[tbp]
\includegraphics[scale=0.3]{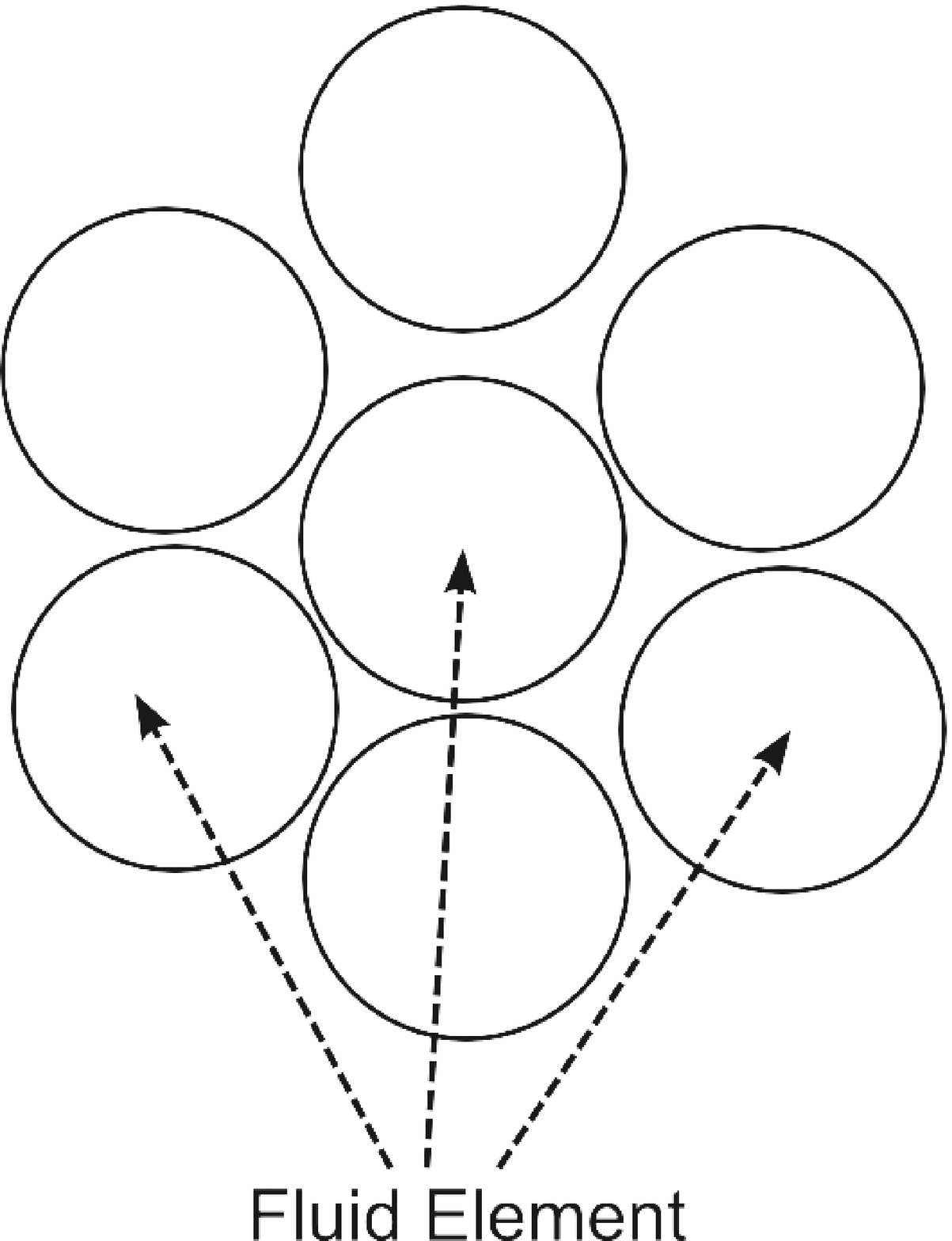}
\hspace{3cm}
\includegraphics[scale=0.3]{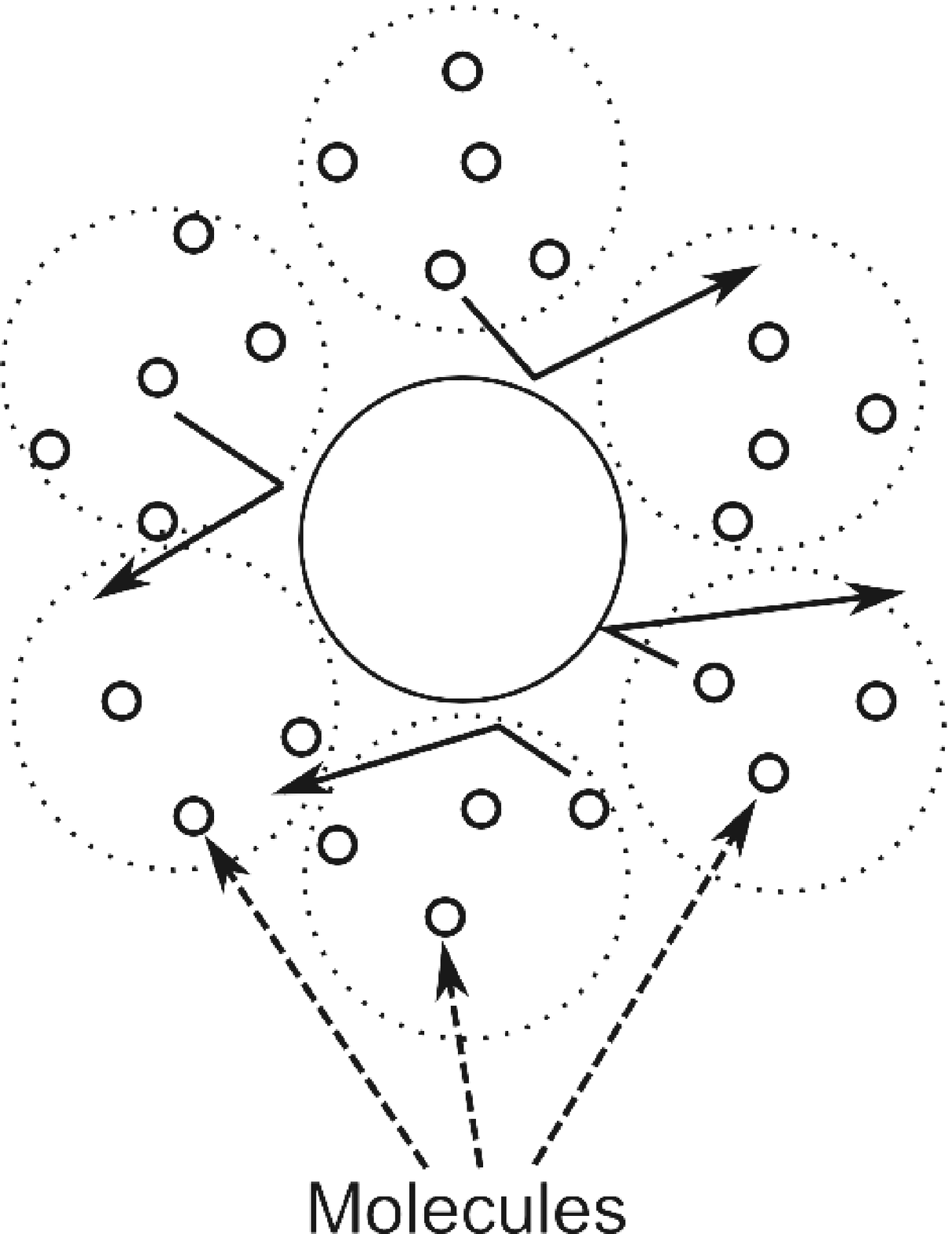}
\caption{The collisions between fluid elements are interpreted as a kind of Brownian motion of a fluid element under the noise of molecules.}
\label{fig1}
\end{figure}

In the stochastic variational method (SVM) \cite{yasue1,morato}, this microscopic effect is taken into account through a different mechanism. Note that a fluid is approximately interpreted as the ensemble of fluid elements inside of which there are many molecules and the thermal equilibrium is achieved. 
The fluid elements move following the flow of the fluid, and the collisions among them are the origin of viscosity. Let us pay attention to the motion of one fluid element as is shown in Fig. \ref{fig1}. Then, because of the fact that each fluid element is consisted from many molecules, the interaction between this and other fluid elements (shown on the left panel) can be regarded as the collisions between this fluid element and molecules (shown on the right panel). This picture is very similar to the Brownian motion.

Motivated by this fact, the definition of velocity of the fluid element is given by a Langevin equation in SVM. Then the information of the microscopic degrees of freedom which is not included in the Lagrangian are introduced through the behavior of the noise. 
In this approach, the fluctuation-dissipation theorem is naturally satisfied and 
the non-dissipative dynamics (the Euler equation in the case of hydrodynamics) is reproduced in the vanishing limit of the noise.

So far, SVM has been exclusively applied to re-formulate the Schr\"{o}dinger equation from the viewpoint of the variational method \cite{qm}. 
However, because of this consistency between SVM and the fluctuation-dissipation theorem, it is natural to expect that SVM is still powerful method to discuss dissipative phenomena.

For the application of SVM to derive the incompressible Navier-Stokes equation has been studied \cite{yasue2}. Recently, it was shown that even the compressible Navier-Stokes equation can be derived \cite{kk1,kk2}. 
In this paper, we generalize the previous argument. We parameterize the stochastic Lagrangian of SVM in more general form and introduce a Hamiltonian 
in terms of the Lagrange transform.

This paper is organized as follows.
In Sec. 2, we develop SVM in the Hamiltonian formulation and apply it to the derivation of the Navier-Stokes equation. In Sec. 3, we discuss that the fluctuation-dissipation theorem is naturally realized in SVM. Section 4 is devoted to concluding remarks.

\section{Application to hydrodynamics}

In the following, we develop the formulation of SVM and derive the Navier-Stokes equation. The technical details of the derivation are mainly following Refs. \cite{kk1,kk2}.

\subsection{Definition of velocity}

In the classical variational method, the velocity of a particle is a continuous function of time. 
On the other hand, as was mentioned in the previous section, the motion of the fluid element fluctuates because of the random collisions with molecules, and the trajectory is determined by the following stochastic differential equation (SDE), 
\begin{equation}
d\mathbf{r}(t)=\mathbf{u}({\bf r}(t),t) dt+\sqrt{2\nu} d\mathbf{W}(t).~~(dt>0),
\label{eqn:sde}
\end{equation}%
where $\nu$ is the strength of the noise and 
\begin{eqnarray}
E\left[ W_{j}(s)-W_{j}(u)\right] &=&0, \\
\hspace{-1cm}E\left[ (W_{j}(s)-W_{j}(u))(W_{k}(s)-W_{k}(u))\right] &=&\delta
^{jk}|s-u|, \\
E\left[ (W_{j}(s)-W_{j}(u))W_{k}(t)\right] &=&0,
\end{eqnarray}%
In SVM, the effect of the microscopic degrees of freedom which is not included in the Lagrangian is taken into account through the noise $d\mathbf{W}(t)$.

\begin{figure}[tbp]
\includegraphics[scale=0.6]{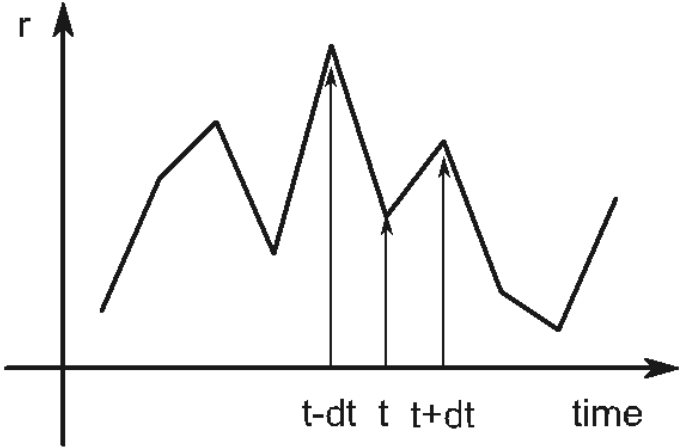}
\caption{The ambiguity for the definition of the time derivative of ${\bf r}$.}
\label{fig2}
\end{figure}

The trajectory ${\bf r}(t)$ is not smooth and hence we cannot define the time derivative at $t$ uniquely, as is shown in Fig. \ref{fig2}. Consequently, we further introduce another SDE which describes the time reversed process of Eq. (\ref{eqn:sde}), 
\begin{equation}
d\mathbf{r}(t)=\tilde{\mathbf{u}}({\bf r}(t),t) dt+\sqrt{2\nu} d\tilde{\mathbf{W}}%
(t),~~(dt<0)  \label{eqn:sde_r}
\end{equation}%
The noise $d\tilde{\mathbf{W}}$ satisfies the same correlation properties as $d{\mathbf{W}}$ and there is no correlation between $d{\mathbf{W}}$ and $d\tilde{\mathbf{W}}$.
The equation must describe the time reversed process of Eq. (\ref{eqn:sde}). To satisfy this condition, the Fokker-Planck equations derived 
from Eqs. (\ref{eqn:sde}) and (\ref{eqn:sde_r}) should be same.
Then $\tilde{\mathbf{u}}$ is related to $\mathbf{u,}$ as 
\begin{equation}
\mathbf{u}=\tilde{\mathbf{u}}+ 2\nu \nabla \ln \rho ,
\label{consist_con}
\end{equation}
where $\rho$ is the particle density which is given by the solution of the Fokker-Planck equation.

Correspondingly to the two SDEs, we introduce the two different definitions of the time derivative \cite{nelson}: 
one is the mean forward derivative, 
\begin{equation}
D {\bf r} = {\bf u},
\end{equation}
and the other the mean backward derivative,
\begin{equation}
\tilde{D} {\bf r} = \tilde{\bf u} .
\end{equation}
The time derivative terms which appear in the kinetic term of the Lagrangian are replaced by 
$D {\bf r}$ and/or $\tilde{D} {\bf r}$.

\subsection{Definition of action}

As is discussed in Ref. \cite{kk1}, the classical action which leads to the Euler equation is given by 
\begin{equation}
I = \int^{t_b}_{t_a} dt \int d^3 {\bf R} 
\rho^{m}_0 \left[ \frac{1}{2} 
\left( \frac{d{\bf r}}{dt} \right)^2 - \frac{\varepsilon}{\rho^m} \right],
\end{equation}
where $t_a$ and $t_b$ are initial and final times.
Note that the action is expressed in the Lagrange coordinate and $\rho^{m}_0$ denotes the initial mass density. The relation between the particle density $\rho$ and the mass density $\rho^m$ is given by $\rho^m/\rho = m$, where $m$ is the mass of the constituent molecules of the fluid. When we apply the classical variational method to this action, we obtain the Euler equation as is discussed in Ref. \cite{kk1}.

As was emphasized in the introduction, the effect of dissipation can be taken into account through the noise in SVM and we still apply the same action for the variation. However, because of the two different definitions of the time derivatives, $D$ and $\tilde{D}$, 
the stochastic representation of this action is not unique.
Note that the dissipative (viscous) terms should disappear and the Euler equation must be reproduced in the vanishing noise limit. 
That is, the stochastic action should be reduced to the classical action in this limit.
Then the most general expression of the stochastic Lagrangian density is given by 
\begin{eqnarray}
I 
&=& 
\int^{t_b}_{t_a} dt \int d^3 {\bf R} E \left[ {\cal L} \right] \label{action_svm} ,
\end{eqnarray}
where the Lagrangian density in the Lagrangian coordinate is 
\begin{eqnarray}
\lefteqn{\cal L=} && \nonumber \\ 
&& \frac{\rho^m_0}{2}\left[ \left( \frac{1}{2} + \alpha_2 \right)\left\{ \left( \frac{1}{2}+\alpha_1 \right) (D{\bf r}(t))^2 
+ \left( \frac{1}{2}-\alpha_1 \right) (\tilde{D}{\bf r}(t))^2 \right\}
+ \left( \frac{1}{2} - \alpha_2 \right) (\tilde{D}{\bf r}(t))(D{\bf r}(t))
 \right] \nonumber \\
&& - \rho^m_0\frac{\varepsilon}{\rho^m} .\label{L}
\end{eqnarray}
Here $\alpha_1$ and $\alpha_2$ is arbitrary constants. 
In this expression, we assume that the terms higher order than $(d{\bf r})^2$ and $(\tilde{D}{\bf r})^2$ do not appear.
One can easily check that, in the vanishing noise limit, $\nu \rightarrow 0$, the difference between $D{\bf r}$ and $\tilde{D}{\bf r}$ disappears, 
\begin{eqnarray}
\lim_{\nu \rightarrow 0} D{\bf r} = \lim_{\nu \rightarrow 0} \tilde{D}{\bf r} = d{\bf r}/dt,
\end{eqnarray}
and hence, independently of the values of $\alpha_1$ and $\alpha_2$, the above action (\ref{action_svm}) always reproduces the classical action.

Note that this action is more general than that used in Refs. \cite{kk1}, which is reproduced by setting $\alpha_1 = 1/2$ and $\alpha_2 = 1/2$.

\subsection{Various formulas}

The most part of the variational procedures in SVM is same as those in the classical variational method.
However we cannot use the Taylor expansion and the partial integration formula. 
These are modified in SVM as follows.
\begin{enumerate}
\item Ito formula \cite{handbook}\\
For an arbitrary function $F$ of the stochastic variable ${\bf r}$ satisfying Eq. (\ref{eqn:sde}), there is the following expansion, 
\begin{equation}
d F({\bf r},t) = \partial_t F + {\bf u}\cdot \nabla F + \nu \nabla^2 F + O(dt^{3/2}).
\end{equation}
\item stochastic partial integration formula \cite{zm}\\
For the stochastic variables ${\bf X}$ and ${\bf Y}$, the partial integration formula is extended as follows,
\begin{equation}
\int_{t_a}^{t_b}dtE[ D{\bf X}(t) \cdot {\bf Y}(t)+ {\bf
X}(t)\cdot \tilde{D}{\bf Y}(t)] 
= E[\mathbf{X}(t_b)\mathbf{Y}(t_b)-\mathbf{X}(t_a)\mathbf{Y}(t_a)].
\end{equation}
\end{enumerate}

\subsection{Stochastic variation}

In Refs. \cite{kk1,kk2}, the stochastic variation was discussed based on the Lagrangian.
Here, we develop the argument using the Hamiltonian which is introduced by the Legendre transform of the Lagrangian.

From Eq. (\ref{L}), the conjugate variables are introduced as 
\begin{eqnarray}
\frac{\rho_0}{2} {\bf p} 
&\equiv& \frac{\partial {\cal L}}{\partial D{\bf r}} 
= \rho^m_0 \left\{ \left( \frac{1}{2} + \alpha_2 \right) \left( \frac{1}{2}+\alpha_1\right) (D{\bf r}) 
+ \frac{1}{2}\left( \frac{1}{2} - \alpha_2 \right)(\tilde{D}{\bf r}) \right\}, \\
\frac{\rho_0}{2}  \bar{\bf p} 
&\equiv& \frac{\partial {\cal L}}{\partial \tilde{D}{\bf r}} 
= \rho^m_0 \left\{ \left( \frac{1}{2} + \alpha_2 \right) \left( \frac{1}{2} - \alpha_1 \right) (\tilde{D}{\bf r}) 
+ \frac{1}{2}\left( \frac{1}{2} - \alpha_2 \right)(D{\bf r}) \right\},
\end{eqnarray}
where the initial particle distribution is given by $\rho_0 = \rho^m_0/m$.
From the Legendre transform, the Hamiltonian corresponding to this Lagrangian is given by 
\begin{eqnarray}
{\cal H}({\bf r},{\bf p},\tilde{\bf p}) 
&\equiv & \frac{\rho_0}{2} \left({\bf p} D{\bf r} + \bar{\bf p} \tilde{D}{\bf r} \right)- {\cal L} \nonumber \\
&=& 
\frac{\rho^m_0}{8 m^2 M}
\left[ \left( \frac{1}{2} + \alpha_2 \right)\left\{ 
\left( \frac{1}{2} -\alpha_1 \right) {\bf p}^2 
+ \left( \frac{1}{2} + \alpha_1 \right) \bar{\bf p}^2 \right\} 
- \left( \frac{1}{2} - \alpha_2 \right) \bar{\bf p}{\bf p}\right] \nonumber \\
&& + \rho^m_0 \frac{\varepsilon}{\rho^m}, 
\end{eqnarray} 
where $M = \alpha_2/2 - \alpha^2_1 \alpha^2_2 - 1/4$. In this derivation, we used that $M \neq 0$.

Then the action which is a function of ${\bf r}$, ${\bf p}$ and $\bar{\bf p}$ is given by 
\begin{eqnarray}
I ({\bf r},{\bf p},\bar{\bf p})
&=& 
\int^{t_b}_{t_a} dt \int d^3 {\bf R} E\left[
\rho_0 \frac{{\bf p} D{\bf r} + \bar{\bf p} \tilde{D}{\bf r}}{2} - {\cal H} 
\right].
\end{eqnarray}
 Note that this action depends on three variables, ${\bf r}$, 
${\bf p}$ and $\bar{\bf p}$.  
Differently from SVM in Refs. \cite{kk1,kk2}, 
we have to consider the variations for these three variables. 
Then we obtain 
\begin{eqnarray}
&& D{\bf r} - \frac{ ( 1/2 + \alpha_2) \left( 1/2 - \alpha_1 \right)}{2mM}{\bf p} + \frac{1/2-\alpha_2}{4mM}\bar{\bf p} = 0, \\
&& \tilde{D}{\bf r} - \frac{( 1/2 + \alpha_2) \left( 1/2 + \alpha_1 \right) }{2mM}\bar{\bf p} + \frac{1/2-\alpha_2}{4mM}{\bf p} = 0, \\
&& \frac{\tilde{D} {\bf p} + D \bar{\bf p}}{2} + m\frac{\delta}{\delta {\bf r}} \frac{\varepsilon}{\rho^m} = 0.
\end{eqnarray}
By eliminating ${\bf p}$ and $\bar{\bf p}$ from the last equation, we obtain 
\begin{eqnarray}
 \left( \frac{1}{2}+\alpha_2 \right) 
\left\{ \left( \frac{1}{2}+\alpha_1 \right) \tilde{D}{\bf u} 
+ \left( \frac{1}{2}-\alpha_1 \right) D\tilde{\bf u} \right\} 
+  \frac{1}{2} \left( \frac{1}{2}-\alpha_2 \right)  ( \tilde{D} \tilde{\bf u} + D{\bf u} ) 
 = - \frac{\delta}{\delta {\bf r}} \frac{\varepsilon}{\rho^m} .
\end{eqnarray}
Here we used Eqs. (\ref{eqn:sde}) and (\ref{eqn:sde_r}) to calculate the forward and backward mean derivatives.

In general, the energy density $\varepsilon$ depends on the specific entropy and hence, to calculate the variation, we have to need certain model for the variation of the entropy. See Refs. \cite{kk1,kk2} for detail.

As a result, the equation corresponds to the Navier-Stokes equation is obtained by 
the stochastic variation, 
\begin{eqnarray}
\rho^m \left( \partial_t + {\bf v}_m \cdot \nabla  \right) {\bf v}^i_m 
+ \partial_j( (P -\zeta \nabla \cdot {\bf v}_m ) \delta^{ij} - \eta e^m_{ij})
-  4\alpha_2 \nu^2 \rho^m  \partial_i ( \sqrt{\rho^m}^{-1} \partial^2_j \sqrt{\rho^m}) 
=0 , \label{NS}
\end{eqnarray}
where $\eta = \alpha_1 ( 1 + 2\alpha_2) \nu \rho^m$, $e^m_{ij}= \partial_j {\bf v}^{i}_m + \partial_i {\bf v}^{j}_m - \frac{2}{3} \delta^{ij} \nabla \cdot {\bf v}_m$ and the fluid velocity ${\bf v}_m$ is given by the mean of ${\bf u}$ and $\tilde{\bf u}$,
\begin{equation}
{\bf v}_m = \frac{{\bf u} + \tilde{\bf u}}{2}.
\end{equation}
In this derivation, we used the model for the variation of entropy used in Ref. \cite{kk1} and then the bulk viscous coefficient $\zeta$ is introduced.
This result is the generalization of that of Ref. \cite{kk1,kk2}.
When we use $\alpha_1 = \alpha_2 = 0$, the effects of the noise cancel each other and 
Eq. (\ref{NS}) is reduced to the Euler equation by ignoring the effect from the variation of the entropy, that is, $\zeta = 0$.
When we choose $\alpha_1 = 0$, $\alpha_2 = 1/2$ and appropriate $\nu$ and energy density $\varepsilon$, we obtain the Gross-Pitaevskii equation.
See Ref. \cite{kk2} for details.

When we set $\alpha_2 = 0$ and abrbitrary finite $\alpha_1$, this equation is completely equivalent to the Navier-Stokes equation.
However, in the framework of SVM, the effect of the noise induces not only viscous terms but also higher order correction term to the 
Navier-Stokes equation.
This correction is not only of second order for the magnitude of fluctuations
(noise) $\nu$, but also of third order for the spatial
derivative $\nabla$. The Navier-Stokes equation does not contain such a term
since, by construction, only the first and second order spatial derivative
terms are maintained. In the case of rarefied gases, this corresponds to the
first order truncation in the derivative (Chapman-Enskog) expansion of
one-particle distribution functions.
However, there is a possibility that this correction term becomes important in the discussion of 
very viscous fluids and the behaviors of the shock wave.
Interestingly, this term may be identified with the surface effect of the Thomas-Fermi approximation used in the nuclear physics.

In SVM, the behavior of the mass density $\rho^m$ is automatically determined by the Fokker-Plank equation corresponding to Eq. (\ref{eqn:sde}) (and (\ref{eqn:sde_r})). By using ${\bf v}_m$, 
this is expressed as
\begin{equation}
\partial_t \rho^m + \nabla (\rho^m {\bf v}_m) = 0 .
\end{equation}
This is the conservation equation of the mass density.

\section{Fluctuation-Dissipation theorem}

As was mentioned in Ref. \cite{kk1}, the noise introduced in
Eq. (\ref{eqn:sde}) is related directly to the transport coefficients. For
example, we can show that the kinetic viscosity $\eta/\rho^m$ satisfies the Einstein relation, 
\begin{equation}
\frac{\eta}{\rho^m} = \frac{1}{3} \alpha_1 ( 1+ 2\alpha_2) \int_{0}^{\infty }dtE[\delta \widehat{\mathbf{v}}%
(t)\cdot \delta \widehat{\mathbf{v}}(0)],  \label{fd2nd}
\end{equation}%
where $\delta \widehat{\mathbf{v}}=d\mathbf{r}/dt-\mathbf{u=}\sqrt{2\nu }d%
\mathbf{W/}dt$. 
Note that the result in Ref. \cite{kk1} is reproduced by using $\alpha_1 = 1/2$ and $\alpha_2 = 1/2$.
This is nothing but the realization of the fluctuation-dissipation theorem 
and it appears as a natural consequence of SVM. We thus conclude that SVM 
possesses not only the well-defined mathematical structure but also a reasonable 
mechanism of dissipation.

\section{Concluding remarks}

We applied the stochastic variational method to the action of the ideal fluid and showed that the Navier-Stokes equation is naturally derived. 
Differently from the previous works \cite{kk1,kk2}, we parameterized the Lagrangian of SVM in more general form. The structure of the obtained equation is not modified but the definition of the transport coefficients were changed.
We further discussed the formulation of SVM using the Hamiltonian and showed that 
the same result can be obtained from the variation of the Hamiltonian.

The stochastic Hamiltonian is discussed also in Ref. \cite{cress}, but only one conjugate momentum is introduced there. 
On the other hand, two conjugate momenta are necessary in our approach. 
The existence of the two different momenta becomes important to discuss the uncertainty relation in hydrodynamics \cite{kk3}.

What is the importance of the formulation of hydrodynamics in the framework of the variational method?
If this approach can be established as another principle of variation, 
we can derive various dissipative equations in more systematic manner.  
As a matter of fact, the variational method with the Rayleigh dissipation function has been used to derive the evolution equations in soft matter physics \cite{doi}. 
And it may become possible to discuss the general properties of different dissipative phenomena 
from the view point of the symmetry of Lagrangian.
As for the Noether theorem in SVM, see Ref. \cite{noe}.

As a matter of fact, in SVM, the effect of dissipation is induced by the effect of fluctuations and the 
fluctuation-dissipation theorem is automatically satisfied. 
This will be an evidence that SVM accounts for the essential mechanism of dissipation, and 
indicates that SVM can be applicable not only to hydrodynamics but also to more general dissipative phenomena where the fluctuation-dissipation theorem is satisfied.

As shown, the present result of SVM specifies the form of the higher order
correction to the Navier-Stokes equation. If SVM is a reliable approach as was discussed above, this higher order term neglected in the Navier-Stokes equation should be considered seriously. 
In fact, as is discussed in Ref. \cite{kk2}, 
this higher order term plays a crucial role in diffusion
processes. The application of SVM to a diffusion process leads to a
generalized form of the diffusion equation which contains memory
effects  \cite{GenDif1}, and the higher order term guarantees that this generalized equation 
reproduce Fick's law in a certain limit.
Moreover, this term becomes significant when the inhomogeneity of the mass density increases. 
Thus the behavior of the shock wave will be affected by this term.

\end{document}